\documentclass[twocolumn]{aastex701}

\usepackage{amsmath}

\begin{document}
\correspondingauthor{Sergio Javier González Manrique}
\shorttitle{Episodic Plasma Flows in an Arch Filament System}
\shortauthors{González Manrique et al.}
\received{2026 June 26}
\revised{2026 September 24}
\accepted{2026 September 25}
\published{2026 October 22}
\journalinfo{Published in ApJL, 1011, L4 (2026); doi:10.3847/2041-8213/aeacdf}
\title{Episodic Plasma Flows and Coronal Brightenings in an Arch Filament System Observed with {\sc Sunrise~iii}}

\author[orcid=0000-0002-6546-5955,sname='González Manrique']{Sergio Javier~González Manrique} \affiliation{Instituto de Astrofísica de Canarias, Vía Láctea, s/n, E-38205 La Laguna, Spain}\affiliation{Departamento de Astrof\'\i sica, Universidad de La Laguna, E-38206 La Laguna, Spain}\email{smanrique@iac.es}

\author[orcid=0000-0001-9857-2573,sname='Trelles Arjona']{Juan Carlos~Trelles Arjona} \affiliation{Instituto de Astrofísica de Canarias, Vía Láctea, s/n, E-38205 La Laguna, Spain}\affiliation{Departamento de Astrof\'\i sica, Universidad de La Laguna, E-38206 La Laguna, Spain}\email{juan.trelles@iac.es}

\author[orcid=0000-0002-3242-1497,sname='Kuckein']{Christoph~Kuckein} \affiliation{Instituto de Astrofísica de Canarias, Vía Láctea, s/n, E-38205 La Laguna, Spain}\affiliation{Departamento de Astrof\'\i sica, Universidad de La Laguna, E-38206 La Laguna, Spain}\affiliation{Spanish Space Solar Physics Consortium}\email{ckuckein@iac.es}

\author[orcid=0000-0003-0175-6232,sname='Siu-Tapia']{Azaymi~L.~Siu-Tapia} \affiliation{Instituto de Astrofísica de Andalucía, CSIC, Glorieta de la Astronomía s/n, 18008 Granada, Spain}\affiliation{Spanish Space Solar Physics Consortium}\email{siu@iaa.es}

\author[sname='Moreno-Insertis']{Fernando~Moreno-Insertis} \affiliation{Instituto de Astrofísica de Canarias, Vía Láctea, s/n, E-38205 La Laguna, Spain}\affiliation{Departamento de Astrof\'\i sica, Universidad de La Laguna, E-38206 La Laguna, Spain}\email{fmi@iac.es}

\author[orcid=0000-0003-1546-381X,sname='Sen']{Samrat~Sen} \affiliation{Instituto de Astrofísica de Canarias, Vía Láctea, s/n, E-38205 La Laguna, Spain}\affiliation{Departamento de Astrof\'\i sica, Universidad de La Laguna, E-38206 La Laguna, Spain}\affiliation{Dr. K. S. Krishnan Geomagnetic Research Laboratory, Indian Institute of Geomagnetism, Prayagraj, India.}\email{samrat.sen@iigm.res.in}

\author[orcid=0000-0002-3418-8449,sname='Solanki']{Sami~K.~Solanki} \affiliation{Max-Planck-Institut für Sonnensystemforschung, Justus-von-Liebig-Weg 3, 37077 Göttingen, Germany}\email{solanki@mps.mpg.de}

\author[orcid=0000-0003-2359-9039,sname='Pastor Yabar']{Adur~Pastor Yabar} \affiliation{Institute for Solar Physics, Department of Astronomy, Stockholm University, AlbaNova University Centre, 10691 Stockholm, Sweden}\email{adur.pastor@astro.su.se}

\author[orcid=0000-0002-1248-0553,sname='Asensio Ramos']{Andrés~Asensio Ramos} \affiliation{Instituto de Astrofísica de Canarias, Vía Láctea, s/n, E-38205 La Laguna, Spain}\affiliation{Departamento de Astrof\'\i sica, Universidad de La Laguna, E-38206 La Laguna, Spain}\email{aasensioramos@gmail.com}

\author[orcid=0000-0002-7788-6482,sname='Nóbrega-Siverio']{Daniel~Nóbrega-Siverio} \affiliation{Instituto de Astrofísica de Canarias, Vía Láctea, s/n, E-38205 La Laguna, Spain}\affiliation{Departamento de Astrof\'\i sica, Universidad de La Laguna, E-38206 La Laguna, Spain}\affiliation{Rosseland Centre for Solar Physics, University of Oslo, PO Box 1029 Blindern, 0315 Oslo, Norway}\affiliation{Institute of Theoretical Astrophysics, University of Oslo, PO Box 1029 Blindern, 0315 Oslo, Norway}\email{dnobrega@iac.es}

\author[orcid=0000-0002-6210-9648,sname='Collados']{Manuel~Collados} \affiliation{Instituto de Astrofísica de Canarias, Vía Láctea, s/n, E-38205 La Laguna, Spain}\affiliation{Departamento de Astrof\'\i sica, Universidad de La Laguna, E-38206 La Laguna, Spain}\email{mcv@iac.es}

\author[orcid=0000-0001-8829-1938,sname='Orozco~Suárez']{David~Orozco~Suárez} \affiliation{Instituto de Astrofísica de Andalucía, CSIC, Glorieta de la Astronomía s/n, 18008 Granada, Spain}\affiliation{Spanish Space Solar Physics Consortium}\email{orozco@iaa.es}

\author[orcid=0009-0007-2532-2371,sname='Rivero González']{María Helena~Rivero González} \affiliation{Instituto de Astrofísica de Canarias, Vía Láctea, s/n, E-38205 La Laguna, Spain}\affiliation{Departamento de Astrof\'\i sica, Universidad de La Laguna, E-38206 La Laguna, Spain}\email{helenariveroglez@gmail.com}

\author[orcid=0000-0002-6807-1273,sname='Esteban Pozuelo']{Sara~Esteban Pozuelo} \affiliation{Instituto de Astrofísica de Canarias, Vía Láctea, s/n, E-38205 La Laguna, Spain}\affiliation{Departamento de Astrof\'\i sica, Universidad de La Laguna, E-38206 La Laguna, Spain}\email{sara.esteban.pozuelo@iac.es}

\author[orcid=0000-0002-3387-026X,sname='del~Toro~Iniesta']{Jose~Carlos~del~Toro~Iniesta} \affiliation{Instituto de Astrofísica de Andalucía, CSIC, Glorieta de la Astronomía s/n, 18008 Granada,   Spain}\affiliation{Spanish Space Solar Physics Consortium}\email{jti@iaa.es}
\author[orcid=0000-0003-4738-7727,sname='Balaguer~Jiménez']{Maria~Balaguer~Jiménez} \affiliation{Instituto de Astrofísica de Andalucía, CSIC, Glorieta de la Astronomía s/n, 18008 Granada, Spain}\affiliation{Spanish Space Solar Physics Consortium}\email{balaguer@iaa.es}
\author[orcid=0000-0002-4208-3575,sname='Sanchis~Kilders']{Esteban~Sanchis~Kilders} \affiliation{Universitat de Valencia Catedrático José Beltrán 2, E-46980 Paterna-Valencia, Spain}\affiliation{Spanish Space Solar Physics Consortium}\email{esteban.sanchis@uv.es}
\author[orcid=0000-0001-9272-6439,sname='Torralbo']{Ignacio~Torralbo} \affiliation{Universidad Politécnica de Madrid,  Plaza Cardenal Cisneros 3, E-28040 Madrid, Spain}\affiliation{Spanish Space Solar Physics Consortium}\email{ignacio.torralbo@upm.es}

\author[orcid=0000-0002-5054-8782,sname='Katsukawa']{Yukio~Katsukawa} \affiliation{National Astronomical Observatory of Japan, 2-21-1 Osawa, Mitaka, Tokyo 181-8588, Japan}\affiliation{Department of Earth and Planetary Science, The University of Tokyo, 7-3-1, Hongo, Bunkyo-ku, Tokyo 113-0033, Japan}\affiliation{Department of Astronomical Science, The Graduate University for Advanced Studies (SOKENDAI), 2-21-1 Osawa, Mitaka, Tokyo 181-8588, Japan}\email{yukio.katsukawa@nao.ac.jp}
\author[orcid=0000-0001-7452-0656,sname='Kawabata']{Yusuke~Kawabata} \affiliation{National Astronomical Observatory of Japan, 2-21-1 Osawa, Mitaka, Tokyo 181-8588, Japan}\email{kawabata.yusuke@nao.ac.jp}
\author[orcid=0000-0001-5686-3081,sname='Hara']{Hirohisa~Hara} \affiliation{National Astronomical Observatory of Japan, 2-21-1 Osawa, Mitaka, Tokyo 181-8588, Japan}\email{hirohisa.hara@nao.ac.jp}
\author[orcid=0000-0003-4764-6856,sname='Shimizu']{Toshifumi~Shimizu} \affiliation{Department of Earth and Planetary Science, The University of Tokyo, 7-3-1, Hongo, Bunkyo-ku, Tokyo 113-0033, Japan}\affiliation{Institute of Space and Astronautical Science, Japan Aerospace Exploration Agency, 3-1-1, Yoshinodai, Chuo-ku, Sagamihara, Kanagawa 252-5210, Japan}\email{shimizu.toshifumi@isas.jaxa.jp}
\author[orcid=0000-0002-7044-6281,sname='Oba']{Takayoshi~Oba} \affiliation{Advanced Research Center for Space Science and Technology, Institute of Science and Engineering, Kanazawa University, Kakuma-machi, Kanazawa, Ishikawa 920-1192, Japan}\affiliation{Max-Planck-Institut für Sonnensystemforschung, Justus-von-Liebig-Weg 3, 37077 Göttingen, Germany}\email{oba@mps.mpg.de}

\author[orcid=0000-0003-1459-7074,sname='Lagg']{Andreas~Lagg} \affiliation{Max-Planck-Institut für Sonnensystemforschung, Justus-von-Liebig-Weg 3, 37077 Göttingen, Germany}\email{lagg@mps.mpg.de}
\author[orcid=0000-0002-9972-9840,sname='Gandorfer']{Achim~Gandorfer} \affiliation{Max-Planck-Institut für Sonnensystemforschung, Justus-von-Liebig-Weg 3, 37077 Göttingen, Germany}\email{gandorfer@mps.mpg.de}
\author[orcid=0000-0002-0787-8954,sname='Bernasconi']{Pietro~Bernasconi} \affiliation{Johns Hopkins University Applied Physics Laboratory, 11100 Johns Hopkins Road, Laurel, Maryland, USA}\email{pietro.bernasconi@jhuapl.edu}
\author[sname='Berkefeld']{Thomas~Berkefeld} \affiliation{Institut für Sonnenphysik (KIS), Georges-Köhler-Allee 401a, 79110 Freiburg, Germany}\email{thomas.berkefeld@leibniz-kis.de}
\author[orcid=0009-0009-4425-599X,sname='Feller']{Alex~Feller} \affiliation{Max-Planck-Institut für Sonnensystemforschung, Justus-von-Liebig-Weg 3, 37077 Göttingen, Germany}\email{feller@mps.mpg.de}
\author[orcid=0000-0001-6317-4380,sname='Riethmüller']{Tino~L.~Riethmüller} \affiliation{Max-Planck-Institut für Sonnensystemforschung, Justus-von-Liebig-Weg 3, 37077 Göttingen, Germany}\email{riethmueller@mps.mpg.de}

\author[orcid=0000-0001-9228-3412,sname='Álvarez-Herrero']{Alberto~Álvarez-Herrero} \affiliation{Instituto Nacional de T\'ecnica Aeroespacial (INTA), Ctra. de Ajalvir, km. 4, E-28850 Torrejón de Ardoz, Spain}\affiliation{Spanish Space Solar Physics Consortium}\email{alvareza@inta.es}
\author[orcid=0000-0001-5616-2808,sname='Kubo']{Masahito~Kubo} \affiliation{National Astronomical Observatory of Japan, 2-21-1 Osawa, Mitaka, Tokyo 181-8588, Japan}\affiliation{Department of Astronomical Science, The Graduate University for Advanced Studies (SOKENDAI), 2-21-1 Osawa, Mitaka, Tokyo 181-8588, Japan}\email{masahito.kubo@nao.ac.jp}
\author[orcid=0000-0003-3490-6532,sname='Smitha']{H.~N.~Smitha} \affiliation{Max-Planck-Institut für Sonnensystemforschung, Justus-von-Liebig-Weg 3, 37077 Göttingen, Germany}\email{narayanamurthy@mps.mpg.de}
\author[sname='Grauf']{Bianca~Grauf} \affiliation{Max-Planck-Institut für Sonnensystemforschung, Justus-von-Liebig-Weg 3, 37077 Göttingen, Germany}\email{grauf@mps.mpg.de}
\author[sname='Carpenter']{Michael~Carpenter} \affiliation{Johns Hopkins University Applied Physics Laboratory, 11100 Johns Hopkins Road, Laurel, Maryland, USA}\email{michael.carpenter@jhuapl.edu}
\author[sname='Bell']{Alexander~Bell} \affiliation{Institut für Sonnenphysik (KIS), Georges-Köhler-Allee 401a, 79110 Freiburg, Germany}\email{albe@leibniz-kis.de}
\author[orcid=0000-0001-7764-6895,sname='Martínez~Pillet']{Valentín~Martínez~Pillet} \affiliation{Instituto de Astrofísica de Canarias, Vía Láctea, s/n, E-38205 La Laguna, Spain}\affiliation{Spanish Space Solar Physics Consortium}\email{vmpillet@iac.es}

\author[orcid=0000-0002-7318-3536,sname='Bailén']{Francisco~Javier~Bailén} \affiliation{Instituto de Astrofísica de Andalucía, CSIC, Glorieta de la Astronomía s/n, 18008 Granada, Spain}\affiliation{Spanish Space Solar Physics Consortium}\email{fbailen@iaa.es}
\author[orcid=0000-0002-2055-441X,sname='Blanco~Rodríguez']{Julian~Blanco~Rodríguez} \affiliation{Universitat de Valencia Catedrático José Beltrán 2, E-46980 Paterna-Valencia, Spain}\affiliation{Spanish Space Solar Physics Consortium}\email{julian.blanco@uv.es}
\author[orcid=0000-0003-4319-2009,sname='Castellanos~Durán']{Juan~Sebastián~Castellanos~Durán} \affiliation{Max-Planck-Institut für Sonnensystemforschung, Justus-von-Liebig-Weg 3, 37077 Göttingen, Germany}\email{castellanos@mps.mpg.de}
\author[orcid=0009-0002-6808-5154,sname='Harnes']{Edvarda~Harnes} \affiliation{Max-Planck-Institut für Sonnensystemforschung, Justus-von-Liebig-Weg 3, 37077 Göttingen, Germany}\email{harnes@mps.mpg.de}
\author[orcid=0000-0001-6029-7529,sname='Hoelken']{Johannes~Hoelken} \affiliation{Max-Planck-Institut für Sonnensystemforschung, Justus-von-Liebig-Weg 3, 37077 Göttingen, Germany}\email{hoelken@mps.mpg.de}
\author[orcid=0000-0003-1409-1145,sname='Iglesias']{Francisco~A.~Iglesias} \affiliation{Max-Planck-Institut für Sonnensystemforschung, Justus-von-Liebig-Weg 3, 37077 Göttingen, Germany}\affiliation{Grupo de Estudios en Heliofísica de Mendoza, CONICET, Universidad de Mendoza, Boulogne sur Mer 683, 5500 Mendoza, Argentina}\email{iglesias@mps.mpg.de}
\author[orcid=0000-0002-4669-5376,sname='Ishikawa']{Ryohtaroh~T.~Ishikawa} \affiliation{National Institute for Fusion Science, 322-6 Oroshi-cho, Toki City 509-5292, Japan}\email{ishikawa.ryohtaro@nifs.ac.jp}
\author[orcid=0000-0002-1043-9944,sname='Matsumoto']{Takuma~Matsumoto} \affiliation{Centre for Integrated Data Science, Institute for Space-Earth Environmental Research, Nagoya University, Furocho, Chikusa-ku, Nagoya, Aichi 464-8601, Japan}\email{takuma.matsumoto@gmail.com}
\author[orcid=0000-0003-1483-4535,sname='Strecker']{Hanna~Strecker} \affiliation{Instituto de Astrofísica de Andalucía, CSIC, Glorieta de la Astronomía s/n, 18008 Granada, Spain}\affiliation{Spanish Space Solar Physics Consortium}\email{streckerh@iaa.es}
\author[orcid=0000-0003-1971-5551,sname='Vukadinović']{Dušan~Vukadinović} \affiliation{Institut für Physik, Universität Graz, Universitätsplatz 5, 8010 Graz, Austria}\affiliation{Max-Planck-Institut für Sonnensystemforschung, Justus-von-Liebig-Weg 3, 37077 Göttingen, Germany}\email{dusan.vukadinovic@uni-graz.at}

\begin{abstract}
Emerging flux regions (EFRs) are key sites of magnetic flux transport and energy release in the solar atmosphere, often hosting arch filament systems (AFSs). We present high-resolution observations of an EFR obtained during the Sunrise~III mission, combining spectropolarimetric measurements from the Tunable Magnetograph (TuMag) and Sunrise Chromospheric Infrared Spectro-Polarimeter (SCIP) with complementary observations from the Solar Dynamics Observatory (SDO). The dataset provides a continuous view of an AFS over nearly 4~hr. Line-of-sight (LOS) velocities derived from the Mg\,I b$_2$ line reveal three episodic plasma-flow events characterized by strong downflows reaching up to $\sim$17~km~s$^{-1}$ at the AFS footpoints and enhanced upflows near the loop tops. Although this pattern agrees with the classical emergence and draining scenario, the strongest upflows and downflows occur nearly simultaneously. Atmospheric Imaging Assembly (AIA) 304 and 171~\AA\ observations reveal transient extreme-UV (EUV) brightenings in the surrounding emerging system; some occur during episodes of enhanced LOS velocities, while others do not, indicating that the temporal correspondence is not systematic. Multiline spectropolarimetric inversions reveal temperatures between $\sim$4000 and $\sim$10,000~K and a smooth variation of the magnetic field inclination along a representative AFS loop. Non-force-free-field extrapolations provide the large-scale magnetic context of the low-lying AFS loops and surrounding coronal structures. Together, the observations reveal a spatially and temporally complex evolution of the emerging system without establishing a direct causal connection between the local plasma-flow enhancements and the EUV brightenings.
\end{abstract}


\section{Introduction}

Emerging flux regions (EFRs) are the main manifestation of the process by which magnetic fields rise from the solar interior into the solar atmosphere, leading to the formation of active regions and driving a wide range of dynamic phenomena \citep{Parker1955,Zwaan1987,Cheung2008,Cheung2014}. As magnetic flux emerges in the form of rising $\Omega$-loops, it leads to the formation of structured magnetic configurations and associated plasma flows in the overlying atmosphere \citep{MartinezGonzalez2010}. In the chromosphere, this process is commonly observed as arch filament systems (AFSs), consisting of bundles of dark fibrils connecting opposite magnetic polarities and tracing the evolution of newly emerging loops \citep{Bruzek1967,Bruzek1969,GonzalezManrique2017a,GonzalezManrique2020}. These structures are highly dynamic, exhibiting upflows at loop tops and strong, often supersonic, downflows at their footpoints, which reflect plasma drainage along magnetic field lines \citep{Solanki2003b,Lagg2004,Lagg2007,Xu2010,GonzalezManrique2018,Sowmya2022}. 
This basic dynamical picture was already established by classical observations of AFSs. In particular, \citet{Chou1988} showed that individual arch filaments are short-lived structures that rise while plasma drains toward their footpoints, and derived characteristic velocities and thermodynamic properties of the cool material. Later, the comprehensive multiwavelength study of \citet{Malherbe1998} combined high-resolution H$\alpha$ spectroscopy, photospheric magnetic measurements, magnetic-field extrapolations, and Yohkoh X-ray observations, showing that short-lived chromospheric arches continuously reform within a longer-lived AFS and coexist with hotter coronal loops. Their analysis also emphasized the complex magnetic connectivity of emerging systems and the still unresolved relation between flux emergence and heating of the overlying atmosphere.

High-resolution observations have further demonstrated that magnetic flux emergence is considerably more complex than the smooth rise of a single large-scale $\Omega$-loop. The Flare Genesis Experiment provided an important step in this direction. \citet{Bernasconi2002} identified moving dipolar features embedded in nearly horizontal emerging fields and interpreted them as U-shaped portions of flux ropes that remained anchored in the photosphere by dense plasma. Building on these observations, \citet{Pariat2004} showed that emerging magnetic structures can develop a serpentine topology composed of successive $\Omega$- and U-shaped segments. In their scenario, the upper portions of undulatory flux tubes emerge through magnetic buoyancy, while reconnection at bald patches and their associated separatrices allows dipped portions of the field to progressively emerge. This picture suggests that AFSs and the overlying coronal loops may develop through the progressive emergence and reconnection of such undulatory magnetic structures rather than through the rise of a single coherent loop system.

Recent high-resolution observations have revealed that AFSs are not only passive tracers of flux emergence, but also sites of intense small-scale activity, including bidirectional brightenings, jets, and signatures of magnetic reconnection \citep{Zhong2019, Chen2024, Joshi2024}. These processes highlight the important role of AFSs in the transfer of mass and energy through the solar atmosphere. Moreover, filaments are often observed in close association with emerging flux and AFSs, either forming along polarity inversion lines or interacting with newly emerged magnetic fields \citep{Kuckein2009,Kuckein2012a,Kuckein2012b,Yelles2012}. These structures can become unstable and erupt, producing strong plasma motions and changes in magnetic connectivity \citep[see, e.g.,][]{Mackay2010,Parenti2014,Kuckein2020,Shen2022}.

Although the basic plasma-flow pattern of AFSs and the complex, fragmented nature of magnetic flux emergence are therefore well established, the temporal connection between individual episodes of enhanced plasma motion within an AFS and transient energy release in the overlying atmosphere remains much less constrained. Previous observations have demonstrated that cool AFS structures can coexist with hotter coronal loops \citep[e.g.,][]{Malherbe1998}, but establishing how short-lived changes in the chromospheric plasma dynamics are related to brightenings at transition-region and coronal temperatures requires sufficiently long and continuous observations. In particular, extended time series are necessary to distinguish isolated fluctuations from recurrent episodes occurring during the evolution of the same emerging system.

In this work, we present high-resolution, multiline observations of an EFR obtained during the {\sc Sunrise~iii} mission \citep{KorpiLagg2025, Solanki2026}, which builds on an earlier version of Sunrise \citep{Barthol2011} that flew successfully in 2009 \citep{Solanki2010} and 2013 \citep{Solanki2017}. The employed data combine spectropolarimetric measurements from the photosphere to the chromosphere with complementary extreme ultraviolet (EUV) observations from the Atmospheric Imaging Assembly \citep[AIA;][]{Lemen2012} onboard the Solar Dynamics Observatory \citep[SDO;][]{Pesnell2012}. The combination of high spatial resolution, multiline diagnostics, and continuous temporal coverage over nearly four hours allows us to follow the evolution of the AFS over a time interval substantially longer than the lifetime of an individual arch filament and to identify recurrent episodes of enhanced plasma activity.

These observations enable us to simultaneously investigate the evolution of the magnetic and thermodynamic properties inferred from spectropolarimetric inversions, together with the plasma motions derived from Doppler measurements. In particular, we use the long Mg\,I b$_2$ time series to follow the LOS velocities in representative loop-top and footpoint regions of the part of the AFS where the strongest dynamical episodes occur. The nearly four-hour sequence reveals three distinct episodes of enhanced plasma flows, while contemporaneous AIA observations provide the higher-atmospheric context of the emerging system and reveal transient EUV brightenings during its evolution. Our aim is therefore not to revisit the well-established presence of upflows at AFS loop tops and downflows at their footpoints, but to characterize the episodic temporal behavior of these flows within the broader multilayer evolution of the emerging region.

This dataset provides a comprehensive and temporally resolved view of flux emergence across multiple atmospheric layers, offering new observational constraints on the dynamics of AFSs and their evolution within a multilayer atmospheric environment.

\section{Observations}
\label{sec:observations}
The observations were obtained on 2024 July 15 with the {\sc Sunrise~iii} balloon-borne solar observatory. The data cover a time interval from 10:08:10 to 14:04:28~UT, corresponding to a total duration of approximately 3.9~hours. The target region is an EFR located in NOAA active region 13753, observed at heliocentric positions corresponding to $\mu \approx 0.96$ at the beginning of the time series and $\mu \approx 0.95$ at the end. The dataset corresponds to the {\sc Sunrise~iii} observing sequence 29\_EMEF (see \citealt{Solanki2026}).

\begin{figure*}
\centering
\includegraphics[width=\textwidth,height=0.75\textheight,keepaspectratio]{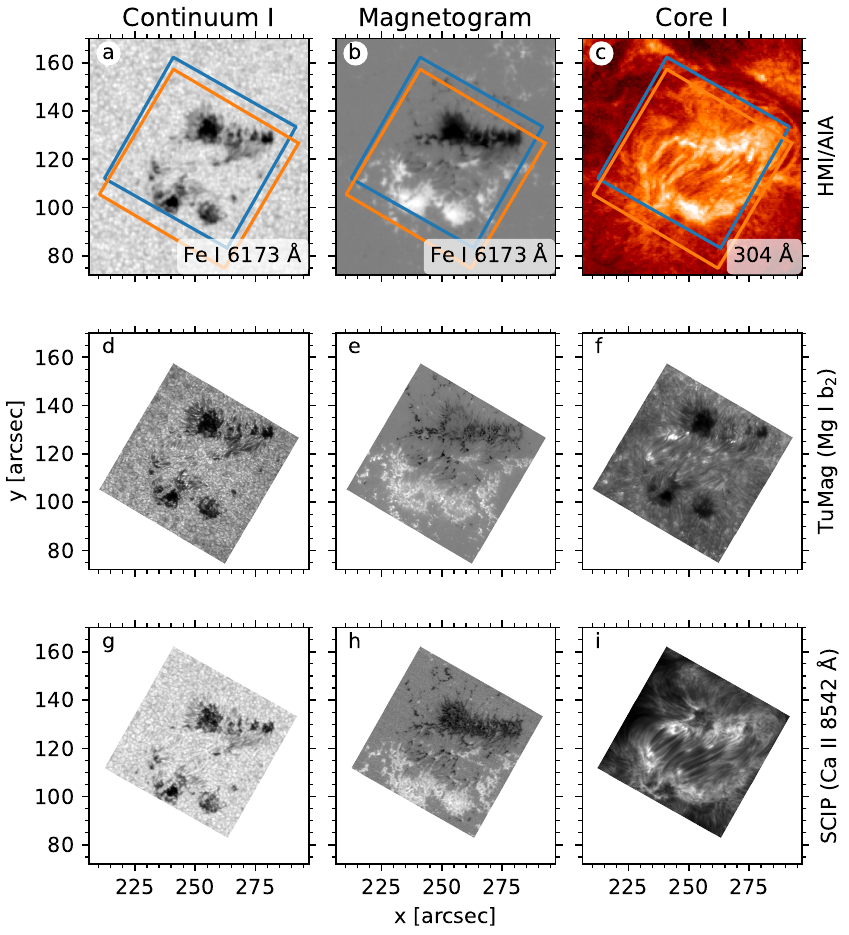}
\caption{
Multi-instrument context of the EFR observed by {\sc Sunrise~iii} and SDO. 
Top row: SDO/HMI continuum intensity ($I_c$), HMI line-of-sight magnetic field ($B_{\mathrm{LOS}}$) derived from the Fe\,I 6173\,\AA\ line, and SDO/AIA 304\,\AA\ image showing the filament and AFS structure. 
The orange and blue boxes indicate the fields of view of the Tunable Magnetograph (TuMag) and Sunrise Chromospheric Infrared Spectro-Polarimeter (SCIP) instruments, respectively. 
Middle row: TuMag reconstructed images in the Mg\,I\,b$_2$ line, including continuum intensity ($I_c$), Stokes $V$ magnetogram derived at 5172.634\,\AA\ in the blue wing of the line, and line-core intensity.
Bottom row: SCIP raster scan in the Ca\,II 8542\,\AA\ line, including continuum intensity ($I_c$), Stokes $V$ magnetogram derived at 8541.135\,\AA\ in the blue wing of the line, and line-core intensity. 
The horizontal and vertical axes in all panels correspond to helioprojective coordinates (arcsec) in the standard WCS reference frame. 
These observations provide the multi-instrument and multiline context of the AFS studied in this work.
An animation of the SDO/HMI and AIA observations is available with the online journal article, covering 10:00:15--14:14:15~UT (the HMI sequence starts at 10:00:45~UT). The five panels show the HMI line-of-sight magnetic field (top left), AIA 304~\AA\ intensity (top center), HMI continuum intensity (top right), AIA 171~\AA\ intensity (bottom left), and AIA 193~\AA\ intensity (bottom center). The observation time is displayed above each panel, and the axes and overlaid coordinate grid indicate helioprojective coordinates in arcsec. The TuMag and SCIP fields of view marked in the static figure are not overplotted in the animation. The sequence shows the evolution of the photospheric magnetic concentrations and sunspots, changes in the morphology and brightness of the surrounding EUV loop structures, and transient EUV brightenings. A partial eruption of the filament north of the main magnetic concentrations is particularly apparent in AIA 304~\AA, beginning around 11:24~UT and followed by changes in the filament morphology.
The animation is available with the \href{https://doi.org/10.3847/2041-8213/aeacdf}{online journal article}.
}
\label{fig:fig1}
\end{figure*}

The region exhibits a complex photospheric structure consisting of sunspots and pores of opposite magnetic polarity, with negative polarity dominating in the northern part and positive polarity in the southern part of the field of view (FOV). In the chromosphere, an AFS connects these opposite polarities. A filament oriented approximately in the east--west direction is observed above the northern part of the active region in AIA (EUV) observations, although it is located outside the FOV of {\sc Sunrise~iii} (see Fig.~\ref{fig:fig1}). 

The dataset captures the evolution of the EFR during a period of filament activity, including the partial eruption of the filament observed in EUV diagnostics, with onset at approximately 11:24~UT, followed by subsequent dynamical changes in the AFS.
{\sc Sunrise~iii} delivers high-resolution, multiline observations of the solar atmosphere with three instruments sampling different atmospheric layers. In this work we make use of data from the Tunable Magnetograph \citep[TuMag;][]{TuMag2025} and the Sunrise Chromospheric Infrared Spectro-Polarimeter \citep[SCIP;][]{SCIP2026} instruments. TuMag provides imaging spectropolarimetric observations in the Fe I 5250.2 \AA\ and the Mg I b$_2$ 5173 \AA\ lines, sampling the photosphere and the lower chromosphere, respectively, while SCIP delivers spectropolarimetric measurements in multiple spectral lines with high spectral resolution. 

For the present study, we focus on the Mg\,I\,b$_2$ observations from TuMag and on the Ca\,II 8542\,\AA\ and Fe\,I 8468\,\AA\ lines from SCIP, which allow us to investigate the atmospheric properties of the arch filament system at different heights in the solar atmosphere. The three lines nicely complement each other, as they cover the photosphere (Fe\,I 8468\,\AA), upper photosphere and lower chromosphere (Mg\,I\,b$_2$), and chromosphere (Ca\,II 8542\,\AA). A detailed study of the formation heights of the former two lines can be found in \citet{Quintero2017}, and of the latter line in \citet{Quintero2018}.

TuMag provides a time series of imaging spectropolarimetric observations in the Mg\,I\,b$_2$ 5173\,\AA\ line over the full duration of the observing sequence. The line is sampled at nine wavelength positions plus one continuum point. The data have an image scale of 0.037$''$\,pixel$^{-1}$, with a diffraction-limited resolution of about 0.13$''$ at the observed wavelengths. A total of 168 maps were acquired with an average cadence of approximately 84~s.
The data were reduced using the TuMag data reduction pipeline\footnote{TuMag's pipeline is publicly available at \url{https://github.com/PabloSGN/TuMags_Reduction_Pipeline}} (D. Orozco Su\'arez et al. 2026, in preparation), which includes a phase-diversity reconstruction module \citep[e.g.,][]{2022ApJS..263....7B}.

All TuMag images within the FOV were first restored using the standard reduction pipeline. The resulting time series was then spatially aligned and derotated using \textsc{torchmfbd} \citep{2025A&A...703A.269A}. The first frame of the sequence was used as a reference, and affine transformations (including translation and rotation) were computed within \textsc{torchmfbd} by maximizing the enhanced correlation coefficient \citep{georgios08} between consecutive frames. The derived transformations were subsequently applied sequentially to the full time series, ensuring a consistent spatial alignment throughout the dataset.

The SCIP instrument repeatedly scanned the full AFS FOV through spectropolarimetric raster scans during the observing sequence. Each map consists of 81 slit positions, covering a FOV of about $58'' \times 58''$, with a cadence of approximately 12~minutes per scan. The spatial sampling is 0.0378$''$ per pixel. SCIP records full Stokes profiles in multiple spectral lines in the near-infrared. The spectral sampling in this wavelength range is approximately 39.5~m\AA\ per pixel.

For the analysis presented here, we use the second raster map, acquired between 11:52:22 and 12:04:17~UT, which corresponds to the end of the filament eruption and to a phase of enhanced dynamics observed in the AFS. Combined with the TuMag observations, these spectropolarimetric data allow us to investigate both the plasma dynamics and the physical properties of the solar atmosphere.

Context observations were obtained from SDO, including data from the Helioseismic and Magnetic Imager (HMI; \citealt{Scherrer2012}) and AIA. HMI provides continuum intensity and LOS magnetograms, which are used to characterize the photospheric structure and magnetic configuration of the region. 

AIA observations at 304, 171, and 193\,\AA\ are used to trace the transition region and coronal structures. The filament and the AFS associated with the EFR are clearly visible in the 304\,\AA\ channel, while the EUV channels reveal the evolution of the overlying plasma during the event. The selected SCIP FOV encompasses the full AFS, although the erupting filament observed in AIA lies outside the SCIP and TuMag FOV. 

An overview of the multi-instrument observations and the relative fields of view is shown in Figure~\ref{fig:fig1}. An animation of the SDO/HMI and AIA observations, covering the time interval from 10:00 to 14:15~UT, is available online.

Complementary full-disk H$\alpha$ observations from GONG were inspected to further characterize the evolution of the filament. 

\section{Analysis}
\label{sec:analysis}

\subsection{Doppler Velocities from Mg\,I\,b$_2$ Profiles}

\begin{figure*}
\centering
\includegraphics[width=\textwidth,height=0.75\textheight,keepaspectratio]{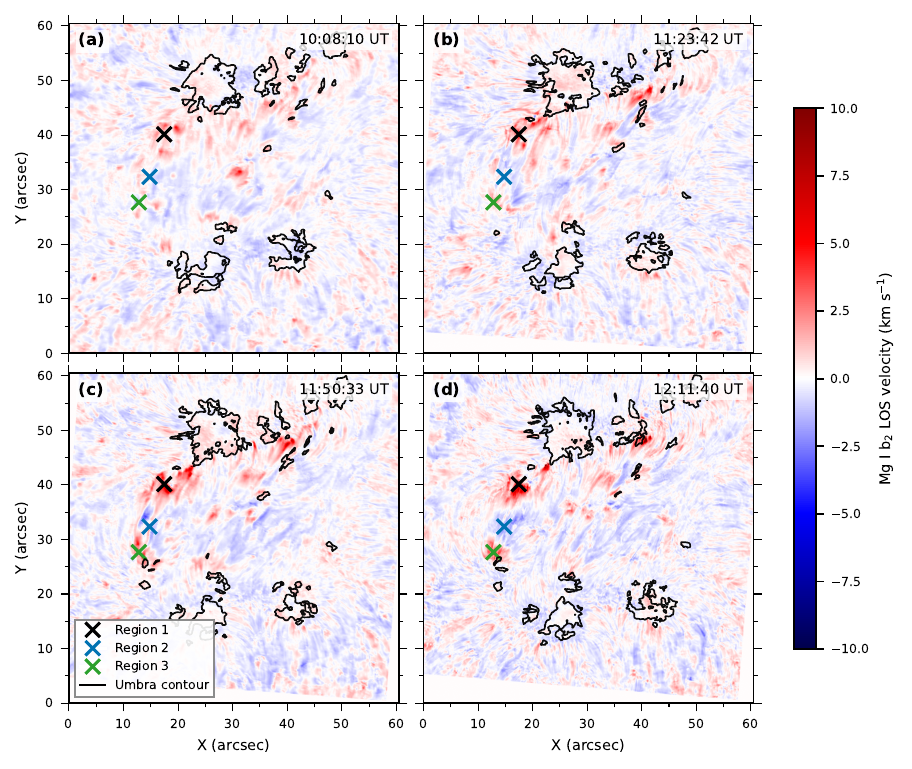}
\caption{
Line-of-sight velocity maps derived from the Mg\,I b$_2$ observations obtained with TuMag during the evolution of the arch filament system. Panels (a)--(d) show representative snapshots at four key stages of the event: (a) the beginning of the Sunrise~III observing sequence (10:08~UT), (b) the onset of the filament eruption (11:23~UT), (c) the end of the eruption phase (11:50~UT), and (d) one of the strongest plasma-flow episodes around 12:11~UT. The crosses mark the centers of the three 3$\times$3 pixel regions used to compute the mean LOS velocities shown in Figure~\ref{fig:fig3}: two footpoints and the loop-top of the arch filament system. These regions were selected within the part of the AFS showing strong plasma activity around 12:11~UT, using the velocity map at 12:11:40~UT (panel d) to identify a representative region of strong loop-top upflows and the corresponding footpoint regions showing enhanced downflows. Once defined, the three regions were ke
pt fixed throughout the time series. Black contours outline the umbral regions derived from the continuum intensity.
An animation of the TuMag observations in the Mg\,I\,b$_2$ line is available with the online journal article, covering 10:08:10--14:03:39~UT on 2024 July 15. The four panels show the continuum intensity (top left), line-core intensity (top right), Stokes~$V$ magnetogram (bottom left), and LOS velocity derived using the Bezier fit (bottom right). The date and time are displayed at the upper left, and the axes give spatial coordinates in arcsec within the approximately 60$\times$60~arcsec field of view. The velocity color scale extends from $-8$ to $+8$~km~s$^{-1}$, with blue and red indicating blueshifts and redshifts, respectively. The crosses marking the three selected regions and the black umbral contours in the static figure are not overplotted in the animation. The sequence shows the evolution of the photospheric intensity and magnetic structures, changes in the dark arch filaments seen in the line core, and the spatially and temporally varying pattern of upflows and downflows within the AFS.
The animation is available with the \href{https://doi.org/10.3847/2041-8213/aeacdf}{online journal article}.
}
\label{fig:fig2}
\end{figure*}

\begin{figure*}
\centering
\includegraphics[width=\textwidth,height=0.80\textheight,keepaspectratio]{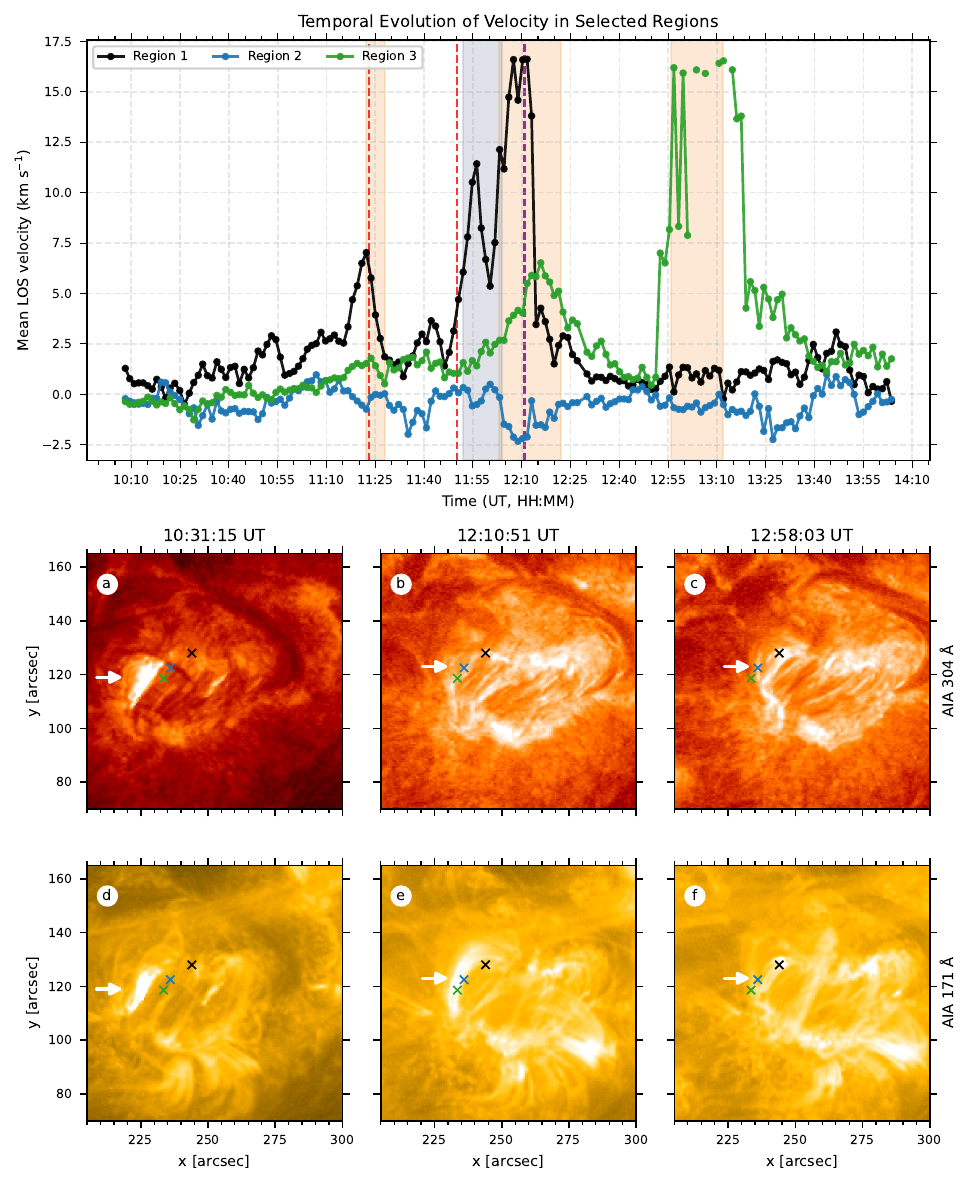}
\caption{
Temporal evolution of the mean LOS velocities in three representative regions of the AFS, together with selected SDO/AIA images illustrating transient EUV brightenings in the surrounding emerging system. The top panel shows the mean LOS velocities derived from the Mg\,I b$_2$ line observed with TuMag. The black and green curves correspond to the two footpoint regions (Regions~1 and 3), while the blue curve represents the loop-top region (Region~2), as indicated in Figure~\ref{fig:fig2}. The vertical dashed red lines mark, respectively, the onset of the filament eruption at 11:24~UT and the end of the eruption phase at 11:50~UT. The gray shaded region indicates the time interval covered by the SCIP raster scan (11:52:22--12:04:17~UT) used for the spectropolarimetric inversions shown in Figure~\ref{fig:fig4}. The violet dashed line marks the time of the representative EUV brightening observed around 12:11~UT. The orange shaded regions indicate the three episodes of enhanced plasma activity identified from the Mg\,I b$_2$ velocity time series: 11:22--11:28~UT, 12:03--12:22~UT, and 12:56--13:12~UT. Panels (a)--(c) show SDO/AIA 304~\AA\ images at 10:31:15, 12:10:51, and 12:58:03~UT, respectively, while panels (d)--(f) show the corresponding SDO/AIA 171~\AA\ images. The white arrows indicate representative transient brightenings. The crosses mark the locations of Regions~1--3 used for the TuMag LOS-velocity analysis. The brightenings at 12:10:51 and 12:58:03~UT occur during the second and third episodes of enhanced LOS velocities, respectively, whereas the brightening at 10:31:15~UT does not coincide with any of the three identified plasma-flow episodes.
}
\label{fig:fig3}
\end{figure*}

LOS velocities are derived from Mg\,I\,b$_2$ spectral profiles observed with TuMag using a Bezier interpolation technique \citep[e.g.,][]{Gordon1974, farin2002}.
For each pixel, a Bezier function is fitted to the spectral samples around the minimum of the Stokes~$I$ profile, and the analytical minimum of the fitted function is used to determine the line-core wavelength position, $\lambda_{\rm lc}$ \citep{SiuTapia2025a,SiuTapia2025b}. The corresponding Doppler velocity is then calculated as $v_{\rm LOS}=c\,\Delta\lambda_{\rm lc}/\lambda_0$, where $\lambda_0=5172.6843$~\AA\ is the reference wavelength of the Mg\,I\,b$_2$ line.
The zero-velocity reference was defined from the average Mg\,I\,b$_2$ profile computed over the full field of view.
Thus, the measured velocities represent Doppler shifts relative to the mean line-core position of the observed field of view. Since the Mg\,I\,b$_2$ line core is mainly sensitive to the upper photosphere and low chromosphere, around the temperature-minimum region, the derived line-core Doppler shifts provide a proxy for plasma motions in these atmospheric layers \citep{SiuTapia2025a,SiuTapia2025b}.
This approach has been successfully applied to Mg\,I\,b$_2$ observations in previous studies \citep{SiuTapia2025a, SiuTapia2025b}, where further details on the implementation can be found. The resulting LOS velocity maps allow us to trace the temporal evolution of the plasma motions within the AFS (see Figure~\ref{fig:fig2} and the bottom-right panel of the animated version of Figure~\ref{fig:fig2}). The temporal evolution of the LOS velocities in the selected regions is shown in Figure~\ref{fig:fig3}.

\subsection{Spectropolarimetric Inversions with DeSIRe}

\begin{figure*}
\centering
\includegraphics[width=\textwidth,height=0.75\textheight,keepaspectratio]{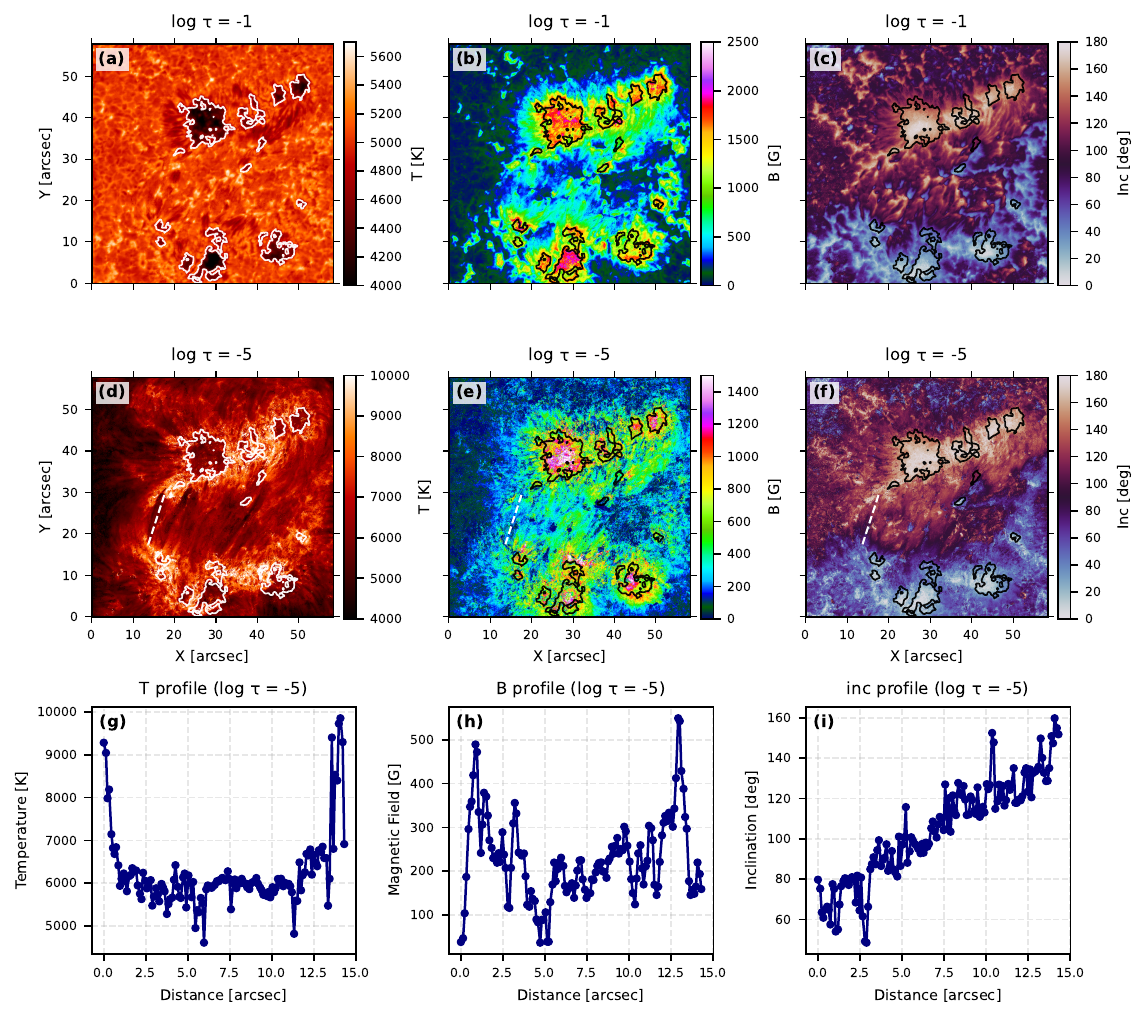}
\caption{
Physical parameters derived from the joint inversion of the Fe\,I 8468\,\AA\ and Ca\,II 8542\,\AA\ lines from SCIP using the DeSIRe inversion code.
The top row shows the inferred temperature, magnetic field strength, and magnetic field inclination at log $\tau = -1$ (photosphere), while the middle row displays the same parameters at log $\tau = -5$ (chromosphere).
Note that the temperature and magnetic field strength maps use different color scales in the first two rows to account for the different thermal and magnetic field regimes of the photosphere and chromosphere.
The black or white contours indicate the umbral regions derived from the continuum intensity.
The white dashed line in panels (d)--(f) traces a representative arch filament, along which the spatial variation of the physical parameters is analyzed to characterize the thermodynamic and magnetic structure of the AFS loop.
The bottom panels (g)--(i) show the corresponding profiles of temperature, magnetic field strength, and magnetic field inclination extracted along this path at log $\tau = -5$ as a function of distance along the arch filament.
For clarity, the magnetic field strength profile shown in panel (h) has been smoothed using a three-point running average.
}
\label{fig:fig4}
\end{figure*}

We performed spectropolarimetric inversions using the DeSIRe code \citep{DeSIRe} on the SCIP dataset. The four Stokes parameters of the \ion{Fe}{1}\,8468\,\AA\ and \ion{Ca}{2}\,8542\,\AA\ lines were inverted simultaneously, providing sensitivity to atmospheric layers from the photosphere to the chromosphere. The inversion reconstructs the atmospheric stratification by perturbing the physical parameters at a discrete set of optical-depth positions (nodes) along the line-of-sight. At each iteration, synthetic Stokes profiles are compared with the observations, and the model atmosphere is updated using the response functions until convergence is achieved. The inversion is performed in successive cycles, progressively increasing the number and distribution of nodes to capture the complexity required by the data. The resulting maps of the inferred atmospheric parameters are shown in Figure~\ref{fig:fig4}. The inversion strategy consisted of four cycles with equally weighted Stokes parameters and a single atmospheric component. Each cycle used the solution from the previous one as the initial guess, while progressively increasing the number of nodes assigned to the atmospheric parameters. This approach allows the inversion to start from a simple atmospheric representation and gradually introduce additional degrees of freedom to refine the fit to the observed Stokes profiles. In the final cycle, up to 11 nodes were used for temperature, 1 for microturbulence, 5 for the magnetic field strength, 11 for the LOS velocity, and 3 nodes each for the inclination and azimuth. The spectral point-spread function (PSF) and spectrally dispersed stray light were determined from the Fe I lines by comparison with the FTS atlas \citep{wallace2003}, following \citet{borrero2016}, yielding an FWHM of 78.59 m\AA{} and a stray-light contribution of 3.4\%. The observations were corrected for this stray light, and the spectral PSF was explicitly included in the inversions; therefore, no macroturbulence was considered. For each pixel, three independent initializations were performed, and the best-fit solution was retained. The initial atmospheric models were based on the semiempirical FAL-C model atmosphere \citep{Fontenla1993}, including small variations in the atmospheric parameters to improve the exploration of the solution space.

\subsection{Non-Force-Free Magnetic Field Extrapolations}

We performed non-force-free field (NFFF) extrapolations to investigate the three-dimensional magnetic configuration of the region at key stages of the event. The method is based on the principle of minimum energy dissipation rate and satisfies the double-curl Beltrami equation \citep{bhattacharyya2007}. In this approach, the magnetic field $\mathbf{B}$ is expressed as the superposition of two linear force-free components and a potential field \citep{Hu2006, Hu2008}. The free parameters of the model are determined by minimizing the deviation between the observed transverse magnetic field and the extrapolated field at the photospheric boundary. 

The extrapolations are computed using HMI vector magnetic field data at two representative times: 11:24~UT, corresponding to the onset of the filament eruption, and 12:12~UT, when enhanced activity is observed in the AFS. Both extrapolations reveal a very similar large-scale magnetic configuration, consistent with the absence of major changes in the HMI photospheric magnetograms during this time interval. In this work, we focus on the magnetic configuration at 11:24~UT, shown in Figure~\ref{fig:fig5}, which illustrates the magnetic structure of the AFS and the filament-associated flux rope. The calculated magnetic field configurations are used to identify regions of complex magnetic connectivity.

\section{Results}
\label{sec:results}

The evolution of the emerging flux region is characterized using context observations from SDO/HMI (continuum intensity and magnetogram) and SDO/AIA 304~\AA\ (Figure~\ref{fig:fig1}), complemented by an online animation showing its temporal evolution. The region exhibits a typical EFR configuration, with opposite-polarity magnetic concentrations connected by an AFS. The AFS is clearly identified as a system of dark loop-like structures visible in the line cores of the Mg\,I\,b$_2$, Ca\,II\,8542\,\AA\ lines and in the SDO/AIA 304 \AA\ image (Figure~\ref{fig:fig1}, panels c, f, and i), tracing cool plasma along magnetic field lines connecting the main polarities. 

A filament located to the north of the main polarities is observed in the AIA 304\,\AA\ channel, showing a dynamic evolution during the observing sequence. The filament initially appears fragmented into two segments that merge shortly before the onset of a partial eruption around 11:24~UT (see online animation). During the eruption phase, which lasts until approximately 11:50~UT, part of the filament disappears, while a remaining portion persists toward the north-west. After the eruption, the filament partially re-forms with a modified morphology. These observations provide a broader context for the evolution of the active region and the AFS, although the filament evolution itself is not analyzed quantitatively in this work.

During this period, episodic brightenings are observed in the AIA 171~\AA\ and 193~\AA\ channels (see online animation), indicating increased activity in the overlying atmosphere. Selected AIA 304~\AA\ and 171~\AA\ images illustrating representative transient brightenings are shown in Figure~\ref{fig:fig3}. These EUV observations are used here primarily to provide context for the evolution of the emerging system and to compare the timing of the brightenings with the episodes of enhanced LOS velocities identified from the TuMag observations. The AFS remains visible throughout the sequence in the AIA 304~\AA, 171~\AA, and 193~\AA\ channels as a system of dark arches connecting the opposite polarities, outlining the evolving magnetic connectivity of the region.

The LOS velocity maps derived from the Mg\,I\,b$_2$ line reveal a clear and structured pattern of plasma motions within the AFS (Figure~\ref{fig:fig2}). The velocity distribution is characterized by upflows in the central parts of the arches, corresponding to the loop-top regions, and downflows concentrated at the footpoints. This pattern is consistent with previous observations of AFSs reported using other chromospheric spectral diagnostics \citep[e.g., ][]{GonzalezManrique2018, Chou1988}. Given the proximity of the observed region to disk center, the LOS velocities mainly trace vertical plasma motions, such that blueshifts and redshifts predominantly correspond to upflows and downflows, respectively. However, the temporal evolution of these plasma motions exhibits pronounced episodic behavior, as discussed below.

Figure~\ref{fig:fig3} shows the temporal evolution of the average LOS velocities measured in the selected 3$\times$3 pixel regions of the AFS. The figure also includes selected AIA 304~\AA\ and 171~\AA\ images at three representative times to provide the broader spatial context of transient EUV brightenings in the emerging system. The regions were selected within the part of the AFS showing strong plasma activity around 12:11~UT, when the Mg\,I b$_2$ velocity map simultaneously shows pronounced upflows near the loop top and strong downflows at the corresponding footpoints. Region 2 samples the location of the strongest loop-top upflows, whereas Regions 1 and 3 sample the two footpoint areas showing enhanced downflows. Each measurement represents the mean velocity over a 3$\times$3 pixel area rather than the velocity of an individual pixel. Once defined, these regions were kept fixed throughout the time series in order to investigate the temporal evolution of the plasma motions at these representative locations. The AIA images are not restricted to these small regions, but are used to illustrate the spatially extended EUV activity surrounding the AFS.

The black and green curves correspond to the two AFS footpoints (Regions 1 and 3), whereas the blue curve represents the loop-top region (Region 2). These fixed regions are not intended to track the same individual arch filament or magnetic field line throughout the observing sequence, but rather to sample the lower-chromospheric plasma dynamics at representative locations within the same part of the emerging system. The time series reveals a sequence of episodic plasma-flow events separated by relatively quiescent periods. During the initial stage, relatively modest velocities are observed, while stronger plasma motions develop later within the AFS, with downflows at the footpoints reaching values of up to $\sim$17~km~s$^{-1}$. Strong upflows are also detected in the loop-top regions and are consistent with the typical plasma motions reported in previous studies of AFSs \citep[e.g.,][]{GonzalezManrique2018,Xu2010}.

Three main episodes of enhanced plasma activity are identified around 11:22--11:28~UT, 12:03--12:22~UT, and 12:56--13:12~UT, corresponding to the orange shaded intervals in Figure~\ref{fig:fig3}. The detection of three separate episodes during the nearly four-hour sequence demonstrates the importance of the extended temporal coverage for distinguishing recurrent plasma-flow enhancements from an isolated event. Interestingly, the strongest downflows are observed only at one footpoint at a time, first in Region~1 and later in Region~3, while enhanced upflows are detected near the loop tops during the same episodes. During the second episode, a small delay is also observed between the maxima of the downflows at the two footpoints, with the strongest downflows in Region~3 occurring a few minutes after those in Region~1.

The selected AIA images shown in Figure~\ref{fig:fig3} illustrate that the temporal correspondence between the transient EUV brightenings and the enhanced LOS velocities is not systematic. At 10:31~UT, a clear brightening is visible in both the 304~\AA\ and 171~\AA\ channels, but it does not coincide with any of the three identified plasma-flow episodes. In contrast, the brightening observed around 12:11~UT, which is particularly prominent in the 171~\AA\ channel, occurs during the second episode of enhanced LOS velocities. A further brightening is observed in both channels around 12:58~UT, coinciding with the beginning of the third plasma-flow episode. Thus, while some EUV brightenings occur contemporaneously with episodes of enhanced chromospheric plasma motions, the observations do not reveal a systematic one-to-one temporal correspondence between the two phenomena.

The temporal evolution does not follow a simple pattern of upflows preceding downflows. Instead, strong upflows and downflows often occur nearly simultaneously, while in some cases the strongest downflows appear before the maximum upflows are reached. These observations reveal a more complex temporal behavior than typically expected in a simple plasma-draining scenario \citep[e.g.,][]{GonzalezManrique2018,GonzalezManrique2020}.

To further investigate the physical structure of the AFS, we analyze the spectropolarimetric inversions derived from the SCIP observations (Figure~\ref{fig:fig4}). The SCIP raster scan was obtained between 11:52:22 and 12:04:17~UT, overlapping the onset of the second major episode of enhanced plasma activity (12:03--12:22~UT) identified in Figure~\ref{fig:fig3}. The simultaneous inversion of the Fe\,I 8468~\AA\ and Ca\,II 8542~\AA\ full-Stokes profiles provides spatially resolved information on the thermodynamic and magnetic properties from the photosphere to the chromosphere. The inferred atmospheric parameters reveal a clear stratified structure along the arch filaments. At chromospheric heights ($\log\tau=-5$), the temperature maps show the AFS loops as elongated cool structures connecting the opposite polarities.

To quantify these variations, we extract the temperature, magnetic field strength, and magnetic field inclination at $\log\tau = -5$ for the pixels along the white dashed line indicated in Figure~\ref{fig:fig4} (panels d--f), which roughly coincides with a representative arch filament. This allows us to exploit the spatial resolution of the SCIP observations to follow the variation of the inferred physical parameters along the same chromospheric structure. The resulting profiles are shown in the bottom panels (g)--(i) of Figure~\ref{fig:fig4} as a function of distance along the arch filament. The selected filament segment lies within the part of the AFS sampled during the second episode of enhanced plasma activity identified in Figure~\ref{fig:fig3}. Thus, the extracted profiles characterize the local thermodynamic and magnetic conditions of the AFS during one of the episodes investigated in the temporal analysis. The temperature profile (Figure~\ref{fig:fig4}(g)) exhibits a clear gradient, with values of $\sim$9000--10,000~K at the footpoints decreasing to $\sim$5000--6000~K toward the loop apex. In some locations along other loops, the temperature drops further to values close to $\sim$4000~K. The magnetic field inclination profile (Figure~\ref{fig:fig4}(i)) varies smoothly along the same structure, transitioning from $\sim$60$^\circ$ at one footpoint to $\sim$160$^\circ$ at the opposite one, consistent with the geometry of a magnetic arch connecting the two polarities. The magnetic field strength profile (Figure~\ref{fig:fig4}(h)) shows larger spatial variations, with values of $\sim$300--400~G at the footpoints and generally lower and more fluctuating values along the loop.

To assess the robustness of these inferred properties, we performed a Monte Carlo experiment for three representative pixels located in the regions corresponding to Footpoint 1, the loop apex, and Footpoint 2. For each pixel, 100 synthetic Stokes profiles were generated by adding Gaussian noise, at the level measured in the observations, to the best-fitting profiles and were subsequently inverted using the same inversion strategy adopted for the observations. The resulting distributions preserve the lower temperature around the loop apex relative to the footpoint regions, as well as the smooth variation of the magnetic field inclination between the two footpoints. The magnetic field strength exhibits comparatively larger relative uncertainties, particularly at $\log\tau=-5$, indicating that its detailed spatial variations should be interpreted more cautiously.

Together, these results provide a spatially resolved view of the thermodynamic and magnetic properties along an AFS loop, directly linking the observed intensity structures with their underlying physical parameters. The derived temperature range is consistent with previous estimates of AFS plasma conditions based on H$\alpha$ observations \citep{Chou1988}, while our analysis additionally resolves the variation of the thermal and magnetic properties along a representative arch filament.

\begin{figure*}
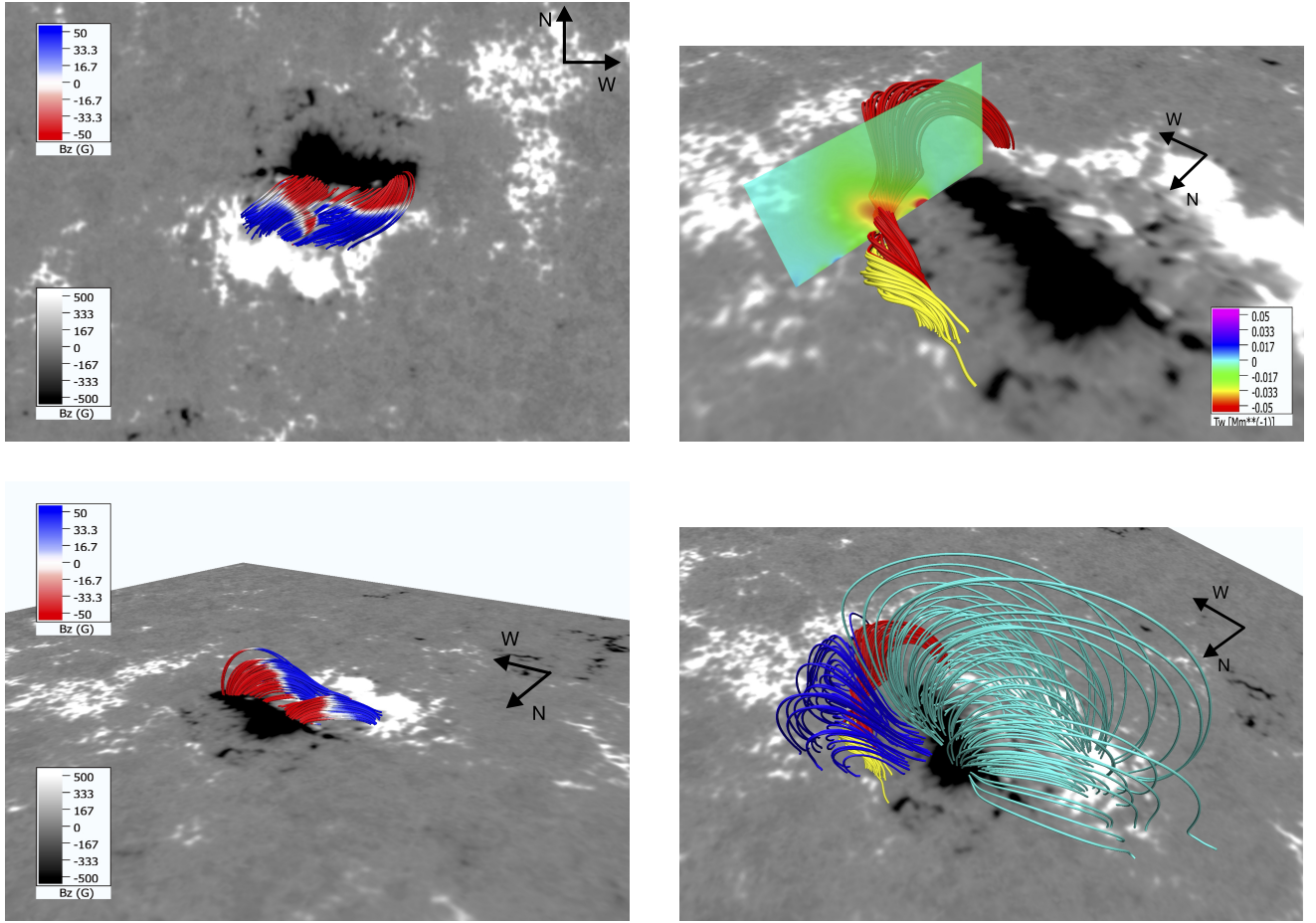

\gridline{
\fig{fig5_1.jpg}{0.46\textwidth}{}
\fig{fig5_3.jpg}{0.46\textwidth}{}
}
\vskip -4mm
\gridline{
\fig{fig5_2.jpg}{0.46\textwidth}{}
\fig{fig5_4_new.jpg}{0.46\textwidth}{}
}
\caption{Illustration of the magnetic configuration at a key stage of the evolution of the EFR (11:24~UT), right before the eruption of the eastern half of the filament. Left column: field lines marking the location of the AFS overlaid on the HMI magnetogram, seen from two perspectives, namely from the side (lower panel) and from the top (upper panel). The latter can be compared with the corresponding images of the AFS in Figure~\ref{fig:fig1}, or Figure~\ref{fig:fig4}, especially panel~(d). The field lines are colored according to the value of $B_z$. The upper-right panel contains field lines, in red and yellow, delineating a twisted flux rope associated with the western side of the filament apparent in Figure~\ref{fig:fig1}, top-right panel; a colormap of the twist density parameter is added on a vertical cut. Lower-right panel: arcade-like flux loops above the stable part of the filament (dark blue) and general AR loops linking the main polarity concentrations of the AR (cyan). The red and yellow field lines of the filament are the same as in the upper-right panel.}
\label{fig:fig5}
\end{figure*}

To gain insight into the magnetic configuration of the region, we analyze the NFFF extrapolations at 11:24~UT, a representative stage of the evolution that coincides with the onset of the partial filament eruption (Figure~\ref{fig:fig5}). The extrapolation is used here to provide the large-scale magnetic context of the AFS and the surrounding active region, rather than to reproduce the fine spatial structure resolved by {\sc Sunrise~iii}. The panels in the left column show sets of magnetic field lines traced from the inner boundary of the major magnetic concentrations as seen from a top and side perspective (upper and lower panels, respectively); the horizontal plane contains a $B_z$-magnetogram. The traced field lines are colored according to the value of $B_z$ on them. In the images one can distinguish two broad groups of field lines, to the East and West of the AR, respectively. The projected geometry of these low-lying field lines closely resembles the arch-like structures observed in the Ca\,II\,8542~\AA\ line-core intensity map shown in Figure~\ref{fig:fig1}, as well as the chromospheric structures visible in Figure~\ref{fig:fig4}. This correspondence supports their identification with the magnetic structures underlying the observed AFS. The maximum height reached by those loops is $\sim 2.5$~Mm and $\sim 5.5$~Mm, for the W and E loop set, respectively.

The upper-right panel shows the magnetic configuration on the N-NW flank of the active region, coinciding with the location occupied by the Western half of the filament, i.e., its stable section, which remains after the eruption of the Eastern half. For better appreciation, this panel can be compared with the top row of Figure~\ref{fig:fig1}. The vertical plane contains a color map with the distribution of the twist density $t_w$, a standard topological quantity defined, e.g., in Eq.~2 of \citet{Moreno-Insertis2025}. There is a patch of high negative twist (yellow-red) at the location of the observed filament. Two sets of field lines are shown as traced from the neighborhood of that patch: the field lines delineate a twisted flux rope, winding anticlockwise around one another with a pitch of roughly $10-20$~Mm along the axial direction, consistent with the value and sign of the twist parameter in the patch. This likely corresponds to the filament channel, with the dense filament material located in its lower levels, at some $2$~Mm above the photosphere, which fits with general configurations expected in filaments \citep{vanBallegooijen1989,Mackay2010,Yelles2012}. Above the flux-rope structure, a system of larger-scale arcade-like loops is present, overlying the stable segment of the filament.

The lower-right panel in Figure~\ref{fig:fig5} illustrates the main magnetic connectivity systems present in the AR and around the filament segment discussed above. The dark blue field lines form arcade-like flux loops located above the stable western portion of the filament, while the cyan field lines correspond to larger-scale active-region loops connecting the main polarity concentrations of the AR. Together with the low-lying AFS loops shown in the left panels, these structures illustrate the multi-scale magnetic environment in which the observed plasma dynamics take place. The system of arcade loops above the filament, like those marked in dark blue in the figure, may help strap and restrain the twisted magnetic rope, thus contributing to the observed stability of that portion of the filament, as abundantly discussed in the literature \citep[e.g.][]{Mackay2010, Gibson2018}.

\section{Discussion}
\label{sec:discussion}

The combined analysis of LOS velocities, spectropolarimetric inversions, and magnetic field extrapolations provides a comprehensive view of the physical processes governing the evolution of the AFS. The observed velocity pattern, characterized by upflows near the loop tops and strong downflows at the footpoints, is broadly consistent with the expected behavior of plasma draining along AFS magnetic loops. This scenario is supported by previous observations of AFSs, where plasma rises to chromospheric heights and subsequently drains toward the lower atmosphere \citep{Chou1988,GonzalezManrique2018,GonzalezManrique2020}. This classical picture is also consistent with the ``leaky bucket'' scenario discussed by \citet{Malherbe1998}, in which individual dark arch filaments empty on short timescales of around 12~min while new loops continuously form, allowing the overall AFS to persist for much longer periods. Similar short-lived and recurrent arch-filament evolution has also been reported in He\,I observations \citep{GonzalezManrique2018,GonzalezManrique2020}.

Our results do not challenge this established plasma-draining scenario. Instead, the long Mg\,I b$_2$ time series allows us to investigate how the lower-chromospheric flows evolve within the same part of the AFS over nearly four hours. The temporal evolution of the LOS velocities reveals a more complex behavior than expected from a simple plasma-draining scenario. The observed plasma motions occur in a sequence of distinct episodic events and are characterized by nearly simultaneous upflows and downflows, suggesting that additional physical processes may contribute to the evolution of the system. The AIA observations provide complementary higher-atmospheric context, showing transient EUV brightenings in the surrounding emerging system. Some of these brightenings occur during episodes of enhanced LOS velocities, while others do not, indicating that the temporal correspondence is not systematic.

The spectropolarimetric inversions provide additional insight into the thermodynamic and magnetic structure of the AFS. The measured temperatures, ranging from $\sim$4000~K to $\sim$10,000~K along the arch filaments, are consistent with previous estimates of AFS plasma conditions based on H$\alpha$ observations \citep{Chou1988}, as well as with the multiwavelength characterization of cool AFS structures and overlying hot loops by \citet{Malherbe1998}, and with more recent studies of AFS dynamics and thermodynamic properties \citep{Chen2024}. In the present observations, the SCIP raster overlapping the onset of the second episode of enhanced plasma activity provides a spatially resolved snapshot of the thermodynamic and magnetic conditions along a representative arch filament. The lowest temperatures are found near the loop apex, while higher temperatures are observed at the footpoints, in agreement with the presence of cool, dense plasma along the arch filament. By exploiting the spatial resolution of the SCIP full-Stokes observations and the sensitivity of the simultaneous Fe\,I and Ca\,II inversions to different atmospheric layers, we characterize the cospatial variation of temperature, magnetic field strength, and magnetic field inclination along the same structure. The smooth variation of the magnetic field inclination along the loop is consistent with a magnetic arch connecting opposite polarities, while the magnetic field strength shows larger spatial variability, with stronger fields concentrated at the footpoints. The Monte Carlo analysis confirms that the lower temperature around the loop apex and the smooth variation of the magnetic field inclination remain robust within the estimated uncertainties, while the larger relative uncertainties in the magnetic field strength require a more cautious interpretation of its detailed spatial variation.

The magnetic extrapolations reveal the large-scale context of these structures, showing that the chromospheric loops inferred from the inversions are embedded within a more extended magnetic configuration reaching coronal heights. The morphology of the low-lying extrapolated field lines is consistent with the arch-like structures observed in the Ca\,II\,8542~\AA\ line-core images, supporting their association with the observed AFS. The inferred heights of the low-lying AFS loops ($\sim 2.5$--$5.5$~Mm) are consistent with previous studies of emerging flux regions, where arch filament systems were found to rise from photospheric heights up to $\sim 8$--$10$~Mm during their evolution \citep{Solanki2003b,GonzalezManrique2020}. Independent height estimates based on Hanle-effect diagnostics also place AFS plasma at heights exceeding $\sim5$~Mm \citep{Merenda2011}, further supporting the inferred loop heights obtained in this work. In addition to these low-lying structures, the extrapolations reveal larger-scale coronal loops connecting more distant polarities of the active region. The resulting coexistence of low-lying AFS loops and overlying coronal magnetic systems is consistent with earlier multiwavelength studies showing cool chromospheric arches beneath hotter coronal loops \citep{Malherbe1998}, and indicates that the observed plasma dynamics take place within a complex multi-scale magnetic environment.

The AFS is embedded in a dynamic magnetic environment that also includes a nearby partial filament eruption. Prior to the eruption, the filament appears fragmented into two segments that merge shortly before the onset of the event, as indicated by additional H$\alpha$ observations from GONG, which are not included in this Letter. Following the eruption, part of the filament disappears, while a remaining portion persists and is cospatial with the magnetic structures identified in the extrapolations. Although the filament eruption occurs close in time to the first episode of enhanced LOS velocities shown in Figure~\ref{fig:fig3} (11:22--11:28~UT), the second and third episodes of enhanced plasma flows occur after the eruption phase has ended. The AIA observations show transient EUV brightenings during the evolution of the emerging system, but their temporal correspondence with the identified plasma-flow episodes is not systematic. Therefore, while some level of interaction between the filament and the AFS cannot be excluded, the available observations do not support a direct causal relationship between the eruption and the full sequence of episodic plasma dynamics discussed above.

The magnetic configuration shown in Figure~\ref{fig:fig5} reveals multiple connectivity systems coexisting within the active region. In the western part of the active region, the extrapolations show a twisted flux rope associated with the stable western portion of the filament discussed above. Above this structure, arcade-like flux loops are present, which may contribute to the stability of the filament material that remains after the partial eruption. More generally, the extrapolations indicate that the low-lying AFS loops are embedded within a complex magnetic environment that includes larger-scale coronal magnetic structures. 

Such a complex configuration is consistent with the picture emerging from the Flare Genesis observations, in which flux emergence proceeds through undulatory or serpentine magnetic structures rather than through the smooth rise of a single coherent $\Omega$-loop. \citet{Bernasconi2002} interpreted moving dipolar features as U-shaped portions of emerging flux ropes still anchored by dense material, while \citet{Pariat2004} showed that successive $\Omega$- and U-shaped segments may emerge progressively, with magnetic reconnection at bald patches and along their separatrices facilitating the rise and reconfiguration of the magnetic field.

The selected AIA images indicate that the temporal relationship between the chromospheric plasma motions and the EUV activity is not systematic. The first episode of enhanced LOS velocities occurs close in time to the onset of the partial filament eruption at 11:24~UT, while the brightenings observed around 12:11 and 12:58~UT occur during the second and third velocity episodes, respectively. It is therefore noteworthy that each of the three episodes of enhanced plasma motions occurs during a period of increased activity in the surrounding system. However, the brightening observed around 10:31~UT does not coincide with any of the three identified plasma-flow episodes, showing that the correspondence is not unique or systematic. The available observations therefore do not allow us to interpret the EUV brightenings as a direct consequence of the strongest chromospheric plasma motions. We use the AIA observations primarily to provide the higher-atmospheric context of the emerging system rather than to establish a direct local connection with the TuMag velocity enhancements.

The episodic nature of the observed plasma flows indicates that the evolution of the AFS is more complex than a simple, continuously evolving pattern of plasma emergence and gravitational draining. The nearly simultaneous occurrence of strong upflows and downflows during the identified episodes suggests that additional intermittent processes may contribute to the dynamics of the system. The AIA observations show that the surrounding higher atmosphere is also highly dynamic, with transient EUV brightenings occurring at different times during the observing sequence. Although some of these brightenings are temporally associated with the identified plasma-flow episodes, the presence of additional brightenings without a corresponding strong velocity enhancement prevents us from establishing a systematic relationship between the two phenomena. The complex magnetic environment inferred from the extrapolations nevertheless provides a natural framework in which different parts of the emerging system may evolve and interact on comparable timescales.

\section{Conclusions}
\label{sec:conclusions}

In this work, we have presented a multi-instrument analysis of an AFS observed with Sunrise~III, combining LOS velocities, spectropolarimetric inversions, and magnetic field extrapolations. The observations reveal a sequence of three episodic plasma-flow events characterized by strong downflows reaching up to $\sim$17~km~s$^{-1}$ at the AFS footpoints and enhanced upflows near the loop tops. The long Mg\,I b$_2$ time series shows that these enhancements recur in the same part of the emerging system, rather than representing the evolution of a single individual arch filament. The AIA 304~\AA\ and 171~\AA\ observations provide complementary higher-atmospheric context, revealing transient EUV brightenings during the evolution of the emerging system. Representative brightenings around 12:11 and 12:58~UT occur during the second and third plasma-flow episodes, respectively, whereas another brightening around 10:31~UT occurs outside the three identified episodes, indicating that the temporal correspondence is not systematic.

The spectropolarimetric inversions provide a spatially resolved characterization of the thermodynamic and magnetic structure along a representative AFS loop, showing temperatures between $\sim$4000 and $\sim$10,000~K, with lower temperatures around the loop apex than at the footpoint regions, and a smooth variation of the magnetic field inclination consistent with an arch connecting opposite polarities. The simultaneous analysis of the Fe\,I and Ca\,II lines reveals the cospatial variation of temperature, magnetic field strength, and magnetic field inclination along the same arch filament, with lower temperatures near the loop apex and higher temperatures toward the footpoints. The Monte Carlo analysis supports the robustness of the temperature and inclination variations, while the detailed spatial changes in magnetic field strength are less tightly constrained and should be interpreted more cautiously. These results extend previous studies by associating the observed chromospheric intensity structure with the inferred local thermodynamic and magnetic properties at the onset of the second episode of enhanced plasma activity.

The comparison between the chromospheric plasma-flow episodes and the AIA observations shows that some transient EUV brightenings occur contemporaneously with enhanced LOS velocities, whereas others do not. Together with the complex magnetic configuration inferred from the extrapolations, these observations indicate that the evolution of the AFS involves multiple spatially and temporally varying components, but they do not establish a direct causal connection between the local plasma-flow enhancements and the EUV brightenings. A dedicated analysis of the spatial morphology, temporal evolution, and magnetic connectivity of the chromospheric and overlying transition-region and coronal structures will be required to investigate this relationship in detail and is beyond the scope of the present Letter. Finally, this study highlights the diagnostic potential of long-duration, multiline, high-resolution observations for investigating the coupling between magnetic field evolution and plasma dynamics in emerging flux regions.

A detailed analysis of the filament eruption and associated jet activity, including their relationship to the surrounding magnetic configuration, is reserved for a separate study based on all available {\sc Sunrise~iii} observations together with complementary spectropolarimetric data from the Swedish Solar Telescope (SST) \citep{Scharmer2003,Scharmer2006} and GREGOR \citep{Schmidt2012,Kleint2020}.

\begin{acknowledgments}
S.J.G.M. acknowledges grant RYC2022-037565-I and C.K. 
grant RYC2022-037660-I, both funded
by MCIN/AEI/10.13039/501100011033 and by “ESF Investing in your future”.
Financial support from grants PID2024-156538NB-I00 and PID2024-156066OB-C55, both funded by MCIN/AEI/
10.13039/501100011033 and by “ERDF A way of making Europe” is gratefully
acknowledged by S.J.G.M. and C.K.. A.L.S.T. acknowledges funding from the Consejer\'ia de Transformaci\'on Econ\'omica, Industria, Conocimiento y Universidades of the Junta de Andaluc\'ia through grant POSTDOC-21-00832.
A.P.Y. acknowledges support from the Swedish Research Council (grant 2023-03313). The Institute for Solar Physics is supported by a grant for research infrastructures of national importance from the Swedish Research Council (registration number 2021-00169). This project has been funded by the European Union through the European Research Council (ERC) under the Horizon Europe program (MAGHEAT, grant agreement 101088184).
This research has been supported by the European Research Council through the
Synergy Grant number 810218 (``The Whole Sun'', ERC-2018-SyG);
by the Research Council of Norway (RCN) through its Centres of Excellence scheme, project number 262622.
S.E.P. acknowledges support from the Agencia Estatal de Investigación del Ministerio de Ciencia, Innovación y Universidades (MCIU/AEI) under grant ``Polarimetric Inference of Magnetic Fields'' and the European Regional Development Fund (ERDF) with reference
PID2022-136563NB-I00/10.13039/501100011033.
{\sc Sunrise~iii} is supported by funding from the Max-Planck-Förderstiftung (Max Planck Foundation), NASA under Grants \#80NSSC18K0934 and \#80NSSC24M0024 (“Heliophysics Low Cost Access to Space” program), and the ISAS/JAXA Small Mission-of-Opportunity program and JSPS KAKENHI Grant Numbers JP18H05234 and JP23K25916. This research has received financial support from the European Union’s Horizon 2020 research and innovation program under grant agreement No. 824135 (SOLARNET) and No. 101097844 (WINSUN) from the European Research Council (ERC). It has also been funded by the Deutsches Zentrum für Luft- und Raumfahrt e.V. (DLR, grant no. 50 OO 1608). The Spanish contributions have been funded by the Spanish MCIN/AEI/10.13039/501100011033 under projects RTI2018-096886-B-C5, PID2021-125325OB-C5, and PID2024-156066OB-C5, and from “Center of Excellence Severo Ochoa” awards to IAA-CSIC (SEV-2017-0709, CEX2021-001131-S), all cofunded by ``ERDF A way of making Europe''. This work is part of grant CEX2025-001609-S, awarded to the Instituto de Astrofísica de Canarias under the Severo Ochoa Centre of Excellence program and funded by MICIU/AEI/10.13039/501100011033. We acknowledge ChatGPT for AI-assisted language editing.
\end{acknowledgments}

\bibliography{biblio}{}
\bibliographystyle{aasjournalv7}

\end{document}